# QCD thermodynamics with Wilson quarks at large $\kappa$

Tom Blum[1], Thomas A. DeGrand[2], Carleton DeTar[3], Steven Gottlieb[4], A. Hasenfratz[2], Leo Kärkkäinen[1], D. Toussaint[1], and R. L. Sugar[5]

(1) Department of Physics, University of Arizona, Tucson, AZ 85721, USA

(2) Physics Department, University of Colorado, Boulder, CO 80309, USA

(3) Physics Department, University of Utah, Salt Lake City, UT 84112, USA

(4) Department of Physics, Indiana University, Bloomington, IN 47405, USA

(5) Department of Physics, University of California, Santa Barbara, CA 93106, USA

(April 12, 1994)

## Abstract

We have extended our study of the high temperature transition with two flavors of Wilson quarks on $12^3 \times 6$ lattices to $\kappa = 0.19$. We have also performed spectrum calculations on $12^3 \times 24$ lattices at $\kappa = 0.19$ to find the physical lattice spacing and quark mass. At this value of $\kappa$ the transition is remarkable in that the plaquette and $\bar\psi\psi$ show a large discontinuity while the Polyakov loop changes very little. This and several other features of the transition are more suggestive of a bulk transition than a transition to a quark-gluon plasma. However, if the temperature is estimated using the $\rho$ mass as a standard, the result is about 150 MeV, in agreement with the value found for the thermal transition with Kogut-Susskind quarks.

12.38.Gc, 11.15.Ha

Typeset using REVTEX

# I. INTRODUCTION

The study of high temperature QCD has proven to be much more difficult with the Wilson formulation of lattice quarks than with the Kogut-Susskind (K-S) formulation. Progress with Wilson quarks has been slow and frustrating. Nevertheless, it is important to continue these studies in order to determine whether the results of lattice simulations are independent of the regularization scheme used. The Wilson formulation provides a way to simulate two flavor QCD without step size errors in the algorithm and without the need to take the square root of the fermionic determinant. The price to be paid is that there is no remnant of the chiral symmetry left to protect the quark mass from additive renormalization, which leads to cumbersome and expensive fine tuning in the search for the chiral limit.

In the first simulations of high temperature QCD with two flavors of Wilson quarks it was found that at the values of $6/g^2$ for which most low temperature calculations were done, $4.5 \leq 6/g^2 \leq 5.7$, the high temperature transition occurs at a value of quark hopping parameter $\kappa$ for which the pion mass measured at zero temperature is quite large [1,2]. In other words, it is difficult to find a set of parameters for which the temperature is the critical temperature and the quark mass is small. Further work confirmed that the pion mass is large at the deconfinement transition for this range of $6/g^2$ [3,4]. (A recent study has concluded that for four time slices the chiral limit is reached at a very small value of $6/g^2 \approx 3.9$ [5].)

Previous simulations with Wilson fermions have located $\kappa_t$, the value of the hopping parameter at which the high temperature crossover or phase transition occurs, as a function of $6/g^2$ for $N_t = 4$ and 6. The critical value of the hopping parameter, $\kappa_c$, for which the pion mass vanishes at zero temperature has been located with somewhat less precision [1,2,6,7,3,4]. Some measurements of hadron masses have been carried out on zero temperature lattices for values of $\kappa$ and $6/g^2$ close to the $\kappa_t$ curve, allowing one to set a scale for the temperature, and to estimate $\kappa_c$ in the vicinity of the thermal transition [3,4,9]. In a recent work with four time slices we have found that the transition or crossover is steepest for $\kappa \approx 0.19$, becoming more gradual for larger or smaller $\kappa$ [10].

In our previous work at $N_t = 6$, we observed coexistence of the low and high temperature phases over long simulation times at $\kappa = 0.17$ and $0.18$. The change in the plaquette across the transition is much larger than for the high temperature transition with Kogut-Susskind quarks [4]. We have extended these observations in the present project. First, we carried out a series of runs on $12^3 \times 6$ lattices at $\kappa = 0.19$. Here our hope was to explore the high temperature transition at a smaller physical quark mass. This would mean a smaller pion to rho mass ratio at the crossover. (For both $N_t = 4$ and 6 the $\pi$ to $\rho$ mass ratio at the thermal crossover decreases as $\kappa$ increases.) We also made hadron spectrum measurements on $12^3 \times 24$ lattices at $\kappa = 0.19$ so that we could find the physical lattice spacing at the transition. We have made a short set of runs at $\kappa = 0.20$ in order to determine the location of the thermal transition for this value of $\kappa$, and have carried out a series of runs on the high temperature side of the transition in the region $0.18 \leq \kappa \leq 0.19$ in order to obtain more information on the nature of the transition.

## II. SIMULATIONS AND RESULTS

| $N_t$ | $\kappa$ | $6/g^2$ | sim. time | ignore | dt | accept |
|---|---|---|---|---|---|---|
| 6 | 0.19 | 4.76 | 100 | 20 | 0.0141 | 0.79 |
| 6 | 0.19 | 4.78 | 120 | 30 | 0.0118 | 0.76 |
| 6 | 0.19 | 4.80c | 520 | 100 | 0.0118 | 0.67 |
| 6 | 0.19 | 4.80h | 480 | 100 | 0.0088 | 0.62 |
| 6 | 0.19 | 4.81 | 595 | 300 | 0.0118 | 0.73 |
| 6 | 0.19 | 4.82 | 715 | 300 | 0.0118 | 0.76 |
| 6 | 0.19 | 4.84 | 990 | 300 | 0.0118 | 0.87 |
| 6 | 0.19 | 4.90 | 970 | 300 | 0.0118 | 0.93 |
| 6 | 0.19 | 4.95 | 1080 | 300 | 0.0118 | 0.94 |
| 6 | 0.20 | 4.54c | 112 | 50 | 0.0071 | 0.42 |
| 6 | 0.20 | 4.54h | 166 | 100 | 0.0071 | 0.64 |
| 6 | 0.1825 | 4.98 | 775 | 300 | 0.0101 | 0.93 |
| 6 | 0.1850 | 4.93 | 1410 | 300 | 0.0101 | 0.88 |
| 6 | 0.1875 | 4.87 | 1286 | 300 | 0.0050 | 0.84 |
| 24 | 0.19 | 4.77 | 728 | 292 | 0.0125 | 0.78 |
| 24 | 0.19 | 4.79 | 688 | 110 | 0.0125 | 0.78 |

TABLE I. Table of new runs. "h" and "c" indicate hot and cold starts.

Simulations were carried out using the hybrid Monte Carlo algorithm with two flavors of dynamical Wilson quarks [8]. The parameters of our new runs are listed in Table I. For the $12^3 \times 6$ lattices we used trajectories with a length of one unit of simulation time in the normalization of Ref. [3]. For the $12^3 \times 24$ lattices we used trajectories one half time unit in length. For reference we show a phase diagram for the relevant range of $\kappa$ and $6/g^2$ in Fig. 1.

In our previous runs with six time slices we found strong metastability at $\kappa = 0.17$ and 0.18. All of the thermodynamic quantities had large discontinuities at these points. However, at $\kappa = 0.19$ the Polyakov loop does not change sharply at the transition, while the plaquette and $\bar{\psi}\psi$ do. This is shown in Fig. 2. Simulation time histories of the Polyakov loop and plaquette for $\kappa = 0.18$ and 0.19 are shown in Figs. 3 and 4 respectively. (Fig. 4 actually contains traces for the Polyakov loop from runs with both hot and cold starts. The values in both runs are small, and are not easily distinguished in this figure.) The short runs at $\kappa = 0.20$ show the same qualitative behavior as the runs at $\kappa = 0.19$.

The different nature of the transition at $\kappa = 0.19$ and 0.18 led us to ask whether there is a sharp change in behavior between these two points, i.e., an intersection of two phase transition lines in the $\kappa, 6/g^2$ plane. We therefore performed runs at $\kappa = 0.1825, 0.1850$ and 0.1875 with $6/g^2$ approximately 0.01 above the transition value — $6/g^2 \approx 6/g_t^2(\kappa)$. The Polyakov loop in these runs, and at $\kappa = 0.19, 6/g^2 = 4.81$ and $\kappa = 0.18, 6/g^2 = 5.02$ with a hot start, are shown in Fig. 5. One sees that the value of the Polyakov loop decreases rather smoothly as $\kappa$ increases.

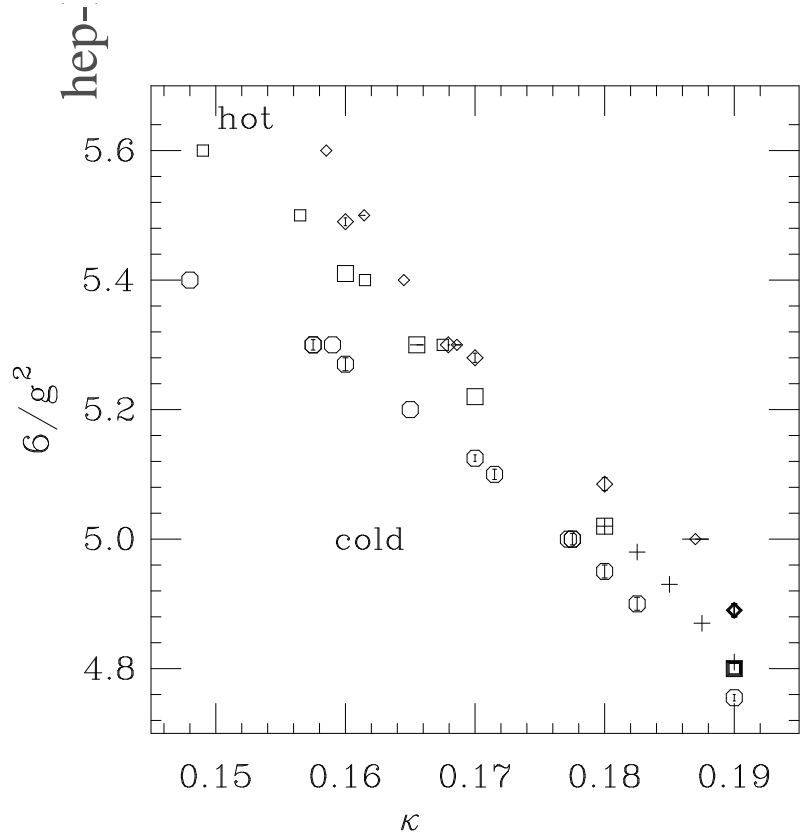

FIG. 1. Phase diagram showing estimates for the high temperature transition and $\kappa_c$. Squares represent the high temperature transition or crossover for $N_t = 6$ and diamonds the zero temperature $\kappa_c$. Octagons are the high temperature crossover for $N_t = 4$. Previous $N_t = 6$ work included in this figure is from Refs. [7] and [4], and the $N_t = 4$ results are from Refs. [7], [3] and [6]. The smaller symbols for $N_t = 6$ and $\kappa_c$ are from older simulations with spatial size eight, and the darker square and diamond are the new results of this work. We show error bars where they are known. For series of runs done at fixed $\kappa$ the error bars are vertical, while for series done at fixed $6/g^2$ the bars are horizontal. The plusses are a set of simulations just on the high temperature side of the $N_t = 6$ thermal transition, which are discussed later (Fig. 5).

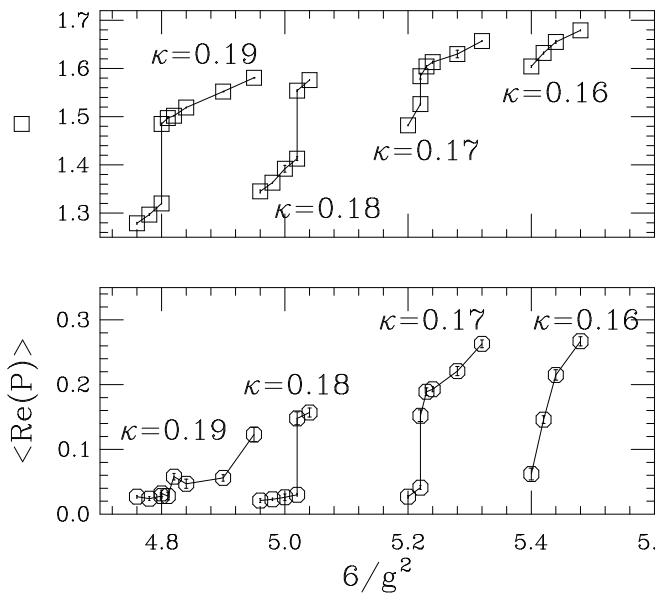

FIG. 2. The plaquette and Polyakov loop versus $6/g^2$ for the different $\kappa$ on lattices with $N_t = 6$.

Following the program of Ref. [4], we made zero temperature runs on $12^3 \times 24$ lattices at $\kappa = 0.19$ to determine the physical quark mass and lattice spacing. It was not practical to run exactly at the transition, $6/g^2 = 4.80$, because the quark matrix was extremely ill-conditioned. Instead, we ran at $6/g^2 = 4.77$ and $4.79$, values for which the simulation was far less costly in cpu time. In addition, a run on an $8^3 \times 16$ lattice at $6/g^2 = 4.76$ was reported in Ref. [3]. The pion and rho masses, and the quark mass from the axial current are tabulated in Table II. From extrapolating the axial current quark mass or the squared pion mass to zero we can estimate the zero temperature gauge coupling, $6/g_c^2$, at which $\kappa_c = 0.19$. Unfortunately, the results from these two quantities do not agree as well as they did at smaller $\kappa$. From extrapolating the quark mass we find $6/g_c^2 = 4.916(16)$, while from extrapolating the squared pion mass we find $6/g_c^2 = 4.886(5)$, with $\chi^2 = 0.6$ for one degree of freedom.

Table III contains the meson masses at the thermal transition. Values for $\kappa = 0.16$, $0.17$ and $0.18$ are from Ref. [4]. The errors in this table include effects of uncertainty in the masses at a particular $6/g^2$ and the effects of an uncertainty of $0.01$ in the crossover value of $6/g^2$ at each $\kappa$. The uncertainty in $6/g_t^2$ is the larger of these effects. At $\kappa = 0.16$ we did not observe metastability, so the crossover value $5.41$ is an estimate of where the physical quantities are changing most rapidly. Because $m_\pi$ and $m_\rho$ both decrease as $\kappa$ increases, the effect of the uncertainty in $6/g_t^2$ on the error in $m_\pi/m_\rho$ is smaller than a naive combination of the errors.

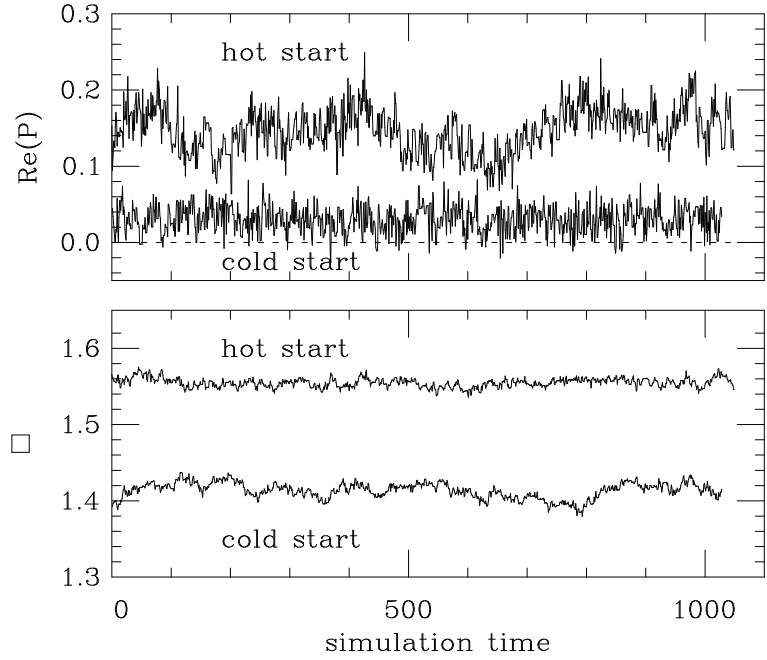

FIG. 3. Coexistence of the plaquette (bottom) and Polyakov loop (top) on a $12^3 \times 6$ lattice at $\kappa = 0.18$ and $6/g^2 = 5.02$ with hot and cold starts.

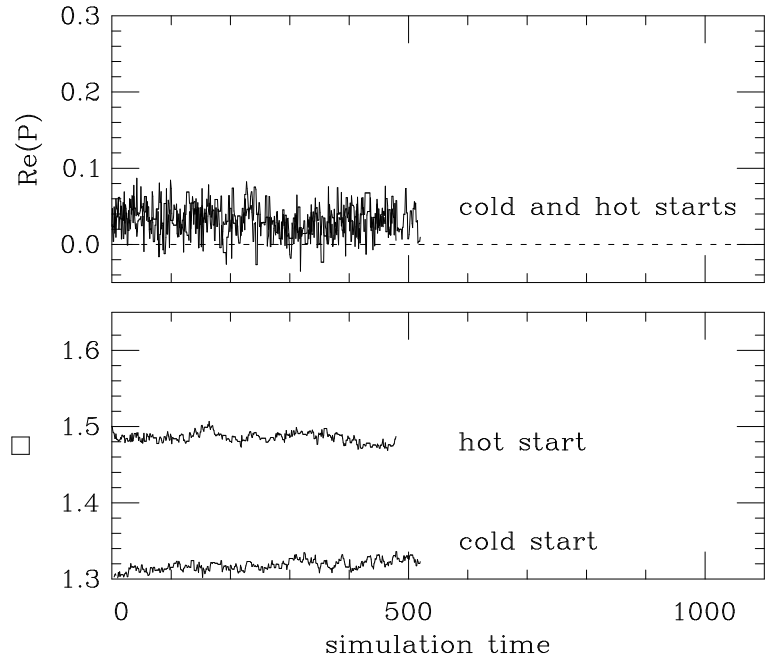

FIG. 4. Coexistence of the plaquette and Polyakov loop on a $12^3 \times 6$ lattice at $\kappa = 0.19$ and $6/g^2 = 4.80$ with hot and cold starts. Here the Polyakov loop is small in both cases.

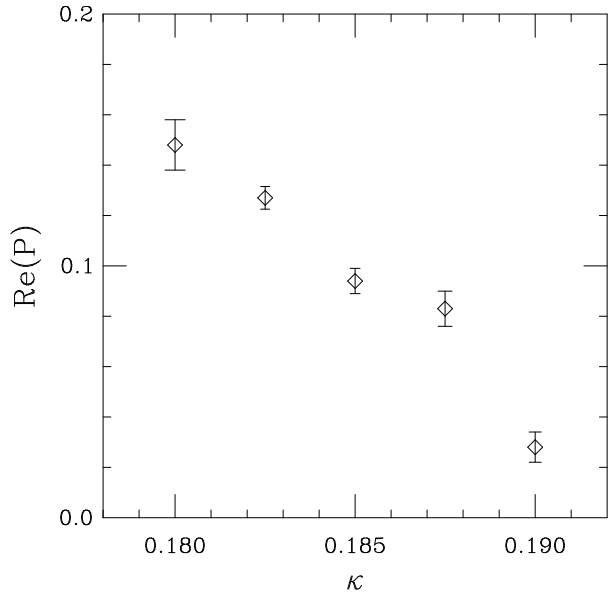

FIG. 5. The Polyakov loop on the high temperature side of the transition. These runs were done at the points marked with plusses in Fig. 1. The point at $\kappa = 0.18$, taken right on the transition $6/g^2$, is from the hot start run (Fig. 3).

| $6/g^2$ | $m_\pi$ | $m_\rho$ | $m_q$ |
|---|---|---|---|
| 4.76 | 0.722(3) | 1.020(9) | NA |
| 4.77 | 0.690(2) | 0.969(7) | 0.0652(7) |
| 4.79 | 0.630(4) | 0.933(7) | 0.0562(8) |

TABLE II. Zero temperature meson masses at $\kappa = 0.19$.

| $\kappa$ | $6/g^2$ | $m_\pi$ | $m_\rho$ | $m_\pi/m_\rho$ |
|---|---|---|---|---|
| 0.16 | 5.41 | 0.73(5) | 0.82(5) | 0.89(1) |
| 0.17 | 5.22 | 0.69(6) | 0.81(6) | 0.85(1) |
| 0.18 | 5.02 | 0.67(6) | 0.89(4) | 0.75(3) |
| 0.19 | 4.80 | 0.60(4) | 0.90(3) | 0.66(3) |

TABLE III. Zero temperature meson masses at the $N_t = 6$ crossover.

## III. DISCUSSION

Because the plaquette and $\bar{\psi}\psi$ are bulk quantities, while the Polyakov loop explicitly tests ordering in the imaginary time, or temperature, direction, it is tempting to speculate that we are seeing a zero temperature transition here. To test this hypothesis by running at larger $N_t$ would be an expensive undertaking.

In Fig. 1 it can be seen that at $\kappa \approx 0.17$ the $N_t = 6$ thermal crossover is well separated from the $N_t = 4$ thermal crossover, and quite close to the estimated $\kappa_c$. However, at $\kappa = 0.19$ the $N_t = 4$ and 6 crossovers are relatively close together, and well separated from $\kappa_c$. This can be dramatized by plotting the zero temperature $m_\pi^2$ at the thermal crossover $6/g^2$, shown in Fig. 6a. An intersection of the $N_t = 4$ and 6 lines on this graph would be equivalent to an intersection of the $N_t = 4$ and 6 lines in Fig. 1. Of course, we could equally well plot $m_\pi^2/m_\rho^2$, shown in Fig. 6b, where an intersection looks much less inevitable. An intersection of the $N_t = 4$ and 6 lines would be characteristic of a bulk transition, as would the vanishing of the gap in the Polyakov loop noted above. It is probably significant that the $N_t = 4$ transition is steepest for $\kappa \approx 0.19$ [10].

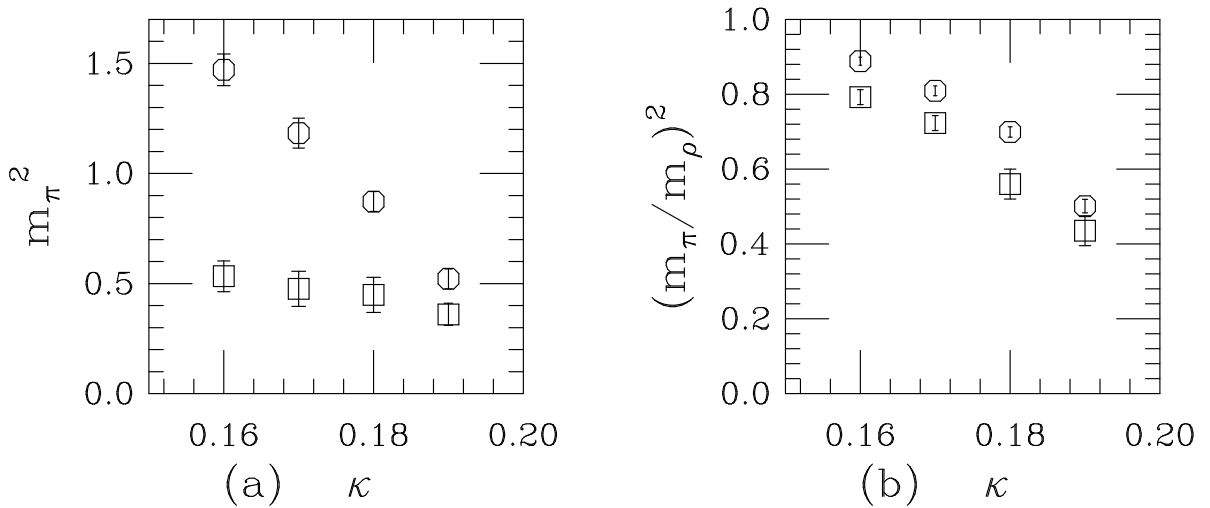

FIG. 6. Pion mass squared as a function of $\kappa$ at the thermal transition line. (a) The squares are for $N_t = 6$ and octagons for $N_t = 4$. The ratio of $(m_\pi/m_\rho)^2$ at $\kappa_t$ as a function of $\kappa$. (b) Most the error comes from the effect of our uncertainty in the location of the thermal crossover, here estimated at $\Delta(6/g^2) = 0.01$.

In Ref. [4] we noted that the change in the plaquette across the $N_t = 6$ thermal transition was surprisingly large. This can be made quantitative from the non-perturbative free energy (or minus one times the pressure) obtained by integrating the plaquette over a range of couplings [13]. With our plaquette normalization,

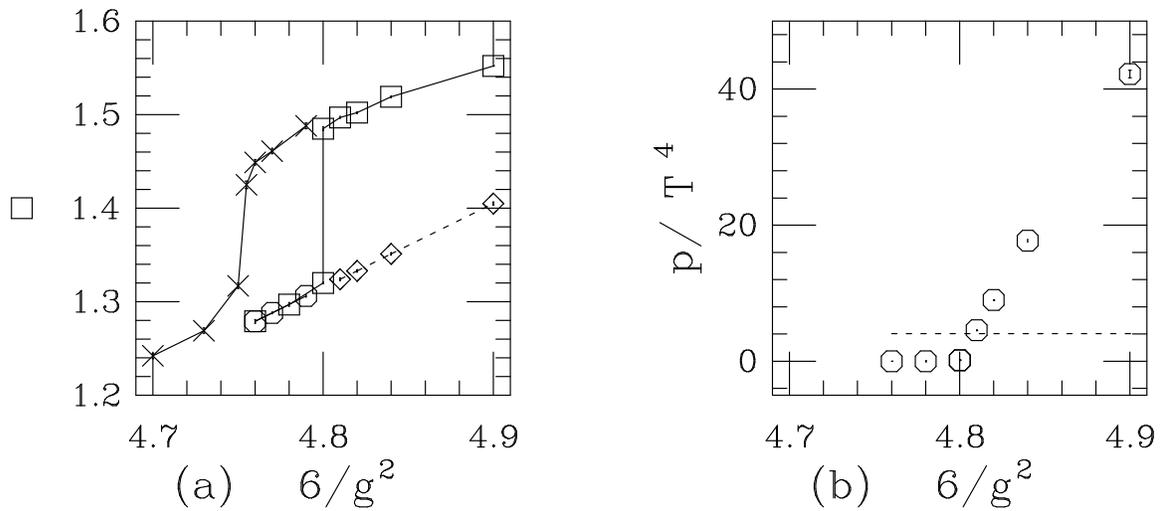

FIG. 7. The $N_t = 4$ (crosses), $N_t = 6$ (squares) and cold (octagons) values of the plaquette as a function of $6/g^2$ at $\kappa = 0.19$. (a) The cold values are extrapolated to larger $6/g^2$ (diamonds) to allow the estimation of the non-perturbative pressure. Non-perturbative $p/T^4$ for $N_t = 6$ as a function $6/g^2$ at $\kappa = 0.19$ assuming the linear extrapolation of the cold plaquettes. (b) The dotted line gives the expected free quark-gluon gas value for two flavors of zero mass quarks.

$$\frac{\partial(pa^4)}{\partial(6/g^2)} = 2\left(\square_{hot} - \square_{cold}\right) \quad . \tag{1}$$

Here $p$ is the pressure and $\square_{hot}$ and $\square_{cold}$ are the average plaquettes on the $12^3 \times 6$ and **extrapolated** $12^3 \times 24$ lattices respectively. The vacuum subtraction is done in the usual way by subtracting the result of the zero temperature simulation. In Fig. 7 we show the plaquettes used in the integration. As discussed above, it was impractical to perform cold runs for $6/g^2 > 4.79$. If we assume that the curve can be linearly extrapolated to somewhat larger $6/g^2$ values, we can still make a crude approximation for the pressure across the transition. The extrapolated values of symmetric plaquettes are shown in Fig. 7 with diamonds. The pressure is then the integral of the difference of the hot and cold plaquettes. Because of the large jump in the plaquette, the integral grows very fast. The Wilson pressure overshoots the ideal gas value by nearly an order of magnitude. Even taking into account the uncertainty in the extrapolation of our cold data points, this clearly demonstrates the huge change of the action in Wilson thermodynamics at large $\kappa$. In the case of Kogut – Susskind fermions, or for smaller $\kappa$ with Wilson fermions, this procedure leads to a pressure comparable to the ideal gas result. In contrast, we note that the change in the plaquette across the transition in Fig. 7a is almost the same for $N_t = 4$ and 6, although there is a clear shift in the position of the transition. Again, this is what one would expect for a bulk transition, while for a thermal transition naive scaling would predict that the gap should decrease by a factor of $(4/6)^4$ in going from $N_t = 4$ to $N_t = 6$.

From previous considerations it is clear that the thermal transition with Wilson quarks is still quite far from giving a trustworthy description of continuum physics. Remarkably, the critical temperature $T_c$ in units of $m_\rho$ is consistent with the K-S quark simulation results when the temporal size is increased to $N_t = 6$. In Fig. 8 we display the current status from K-S studies and our Wilson simulations. While the Wilson results with $N_t = 4$ disagree with K-S estimates, and offer little help in extrapolating to the physical regime, the $N_t = 6$ results nicely line up with those using K-S quarks.

## IV. CONCLUSIONS

We have extended our work on Wilson thermodynamics for $\kappa = 0.19$, finding disturbing and unexpected properties: the lack of a jump in the Polyakov loop while the other indicators of the transition have a clear signal, and the overshooting of the non-perturbative pressure. The lack of a jump in the Polyakov loop, the similarity in the jump in the plaquette at $N_t = 4$ and $N_t = 6$, and perhaps even our inability to perform spectrum calculations at $6/g^2 > 4.79$ are more characteristic of a bulk transition than a transition to a quark gluon plasma. Still, when the temperature of the crossover is estimated in a standard way the results are consistent with $N_t = 6$ work at smaller $\kappa$ and with results using Kogut-Susskind quarks.

Much work remains to be done. At present, the Wilson and Kogut-Susskind formulations do not give us a concise and consistent picture of many aspects of the transition. We hope that future simulations provide the answers to these questions.

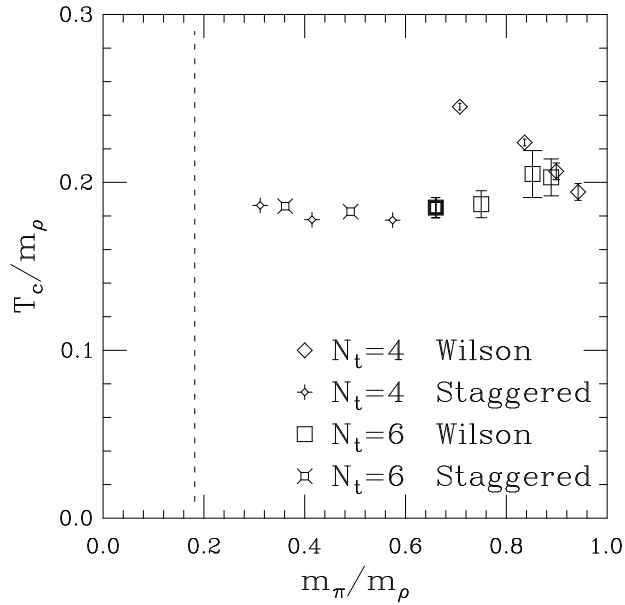

FIG. 8. Critical temperature in units of $m_\rho$ for Wilson and K-S simulations. The darker square is the point added by this work; the remainder of the graph is from Ref. [4]. The dotted line is the physical value of $m_\pi/m_\rho$.

## ACKNOWLEDGMENTS

These calculations were carried out on the iPSC/860 and Paragons at the San Diego Supercomputer Center and at Oak Ridge National Laboratory, and on the CM5 at the National Center for Supercomputing Applications. We are grateful to the staffs of these centers for their help. We also thank Tony Anderson and Reshma Lal of Intel Scientific Computers for their help with the Paragon. We would like to thank Akira Ukawa and Frithjof Karsch for helpful discussions. Several of the authors have enjoyed the hospitality of the Institute for Nuclear Theory, the Institute for Theoretical Physics and the UCSB Physics department, where parts of this work were done. This research was supported in part by Department of Energy grants DE-2FG02–91ER–40628 , DE-AC02–84ER–40125, DE-AC02–86ER–40253, DE-FG02–85ER–40213, DE-FG03–90ER–40546, DE-FG02–91ER–40661, and National Science Foundation grants NSF–PHY93–09458, NSF–PHY91–16964, NSF–PHY89–04035 and NSF–PHY91–01853. L.K. wishes to thank the Antti Wihuri foundation for additional support.